\documentclass[sigconf]{acmart}
\AtBeginDocument{%
  }

\copyrightyear{2026}
\acmYear{2026}
\setcopyright{cc}
\setcctype{by}
\acmConference[SCA/HPCAsia 2026]{SCA/HPCAsia 2026: Supercomputing Asia and International Conference on High Performance Computing in Asia Pacific Region}{January 26--29, 2026}{Osaka, Japan}
\acmBooktitle{SCA/HPCAsia 2026: Supercomputing Asia and International Conference on High Performance Computing in Asia Pacific Region (SCA/HPCAsia 2026), January 26--29, 2026, Osaka, Japan}
\acmPrice{}
\acmDOI{10.1145/3773656.3773670}
\acmISBN{979-8-4007-2067-3/2026/01}

\usepackage[dvipsnames,svgnames]{xcolor}

\usepackage[english]{babel}
\usepackage{microtype}
\usepackage{tikz}
\usetikzlibrary{calc}
\usepackage{pgfplots}
\pgfplotsset{compat=newest}
\usepackage{pgfplotstable}
\usepackage{subcaption}  
\usepackage{caption}
\usepackage{amsfonts}
\usepackage{algorithm}
\usepackage{algpseudocode}

\usetikzlibrary{arrows.meta,positioning,shapes.geometric}

\tikzset{
  fc/line/.style       = {-Latex, thick},
  fc/block/.style      = {rectangle, rounded corners=2pt, draw, align=center,
                          minimum width=3.5cm, minimum height=1.1cm},
  fc/decision/.style   = {diamond, aspect=2, draw, align=center, inner sep=1.5pt},
  fc/term/.style       = {ellipse, draw, align=center, minimum width=2.8cm, minimum height=1.1cm}
}

\usepackage{comment} 

\begin{document}

\title{Guaranteed DGEMM Accuracy While Using Reduced Precision Tensor Cores Through Extensions of the Ozaki Scheme}



\author{Angelika Schwarz}
\affiliation{%
  \institution{NVIDIA Corporation}
  \city{Stockholm}
  \state{}
  \country{Sweden}
}
\email{abschwarz@nvidia.com}

\author{Anton Anders}
\affiliation{%
  \institution{NVIDIA Corporation}
  \city{Santa Clara}
  \state{California}
  \country{USA}
}
\email{antona@nvidia.com}

\author{Cole Brower}
\affiliation{%
  \institution{NVIDIA Corporation}
  \city{Santa Clara}
  \state{California}
  \country{USA}
}
\email{cbrower@nvidia.com}

\author{Harun Bayraktar}
\affiliation{%
  \institution{NVIDIA Corporation}
  \city{Santa Clara}
  \state{California}
  \country{USA}
}
\email{hbayraktar@nvidia.com}

\author{John Gunnels}
\affiliation{%
  \institution{NVIDIA Corporation}
  \city{Santa Clara}
  \state{California}
  \country{USA}
}
\email{jgunnels@nvidia.com}

\author{Kate Clark}
\affiliation{%
  \institution{NVIDIA Corporation}
  \city{Santa Clara}
  \state{California}
  \country{USA}
}
\email{mclark@nvidia.com}

\author{RuQing G. Xu}
\affiliation{%
  \institution{NVIDIA Corporation}
  \city{Minato}
  \state{Tokyo}
  \country{Japan}
}
\email{ruqingx@nvidia.com}

\author{Samuel Rodriguez}
\affiliation{%
  \institution{NVIDIA Corporation}
  \city{Santa Clara}
  \state{California}
  \country{USA}
}
\email{srodriguezbe@nvidia.com}

\author{Sebastien Cayrols}
\affiliation{%
  \institution{NVIDIA Corporation}
  \city{Knoxville}
  \state{Tennessee}
  \country{USA}
}
\email{scayrols@nvidia.com}

\author{Paweł Tabaszewski}
\affiliation{%
  \institution{NVIDIA Corporation}
  \city{Warsaw}
  \country{Poland}
}
\email{ptabaszewski@nvidia.com}

\author{Victor Podlozhnyuk}
\affiliation{%
  \institution{NVIDIA Corporation}
  \city{Reading}
  \country{United Kingdom}
}
\email{vpodlozhnyuk@nvidia.com}



\begin{abstract}
  The rapid growth of artificial intelligence (AI) has made low-precision formats such as FP16, FP8, and, most recently, block-scaled FP4 the primary focus of modern GPUs, where Tensor Cores now deliver orders-of-magnitude higher throughput than traditional FP64 pipelines. This hardware shift has sparked a new line of algorithm research: using low-precision units to emulate double-precision accuracy through schemes such as Ozaki decompositions. We advance this direction with Automatic Dynamic Precision (ADP), a fully GPU-resident framework that makes emulated FP64 matrix multiplication both efficient and reliable. At its core is the Exponent Span Capacity (ESC), a hardware-agnostic estimator that conservatively determines the decomposition parameter (a.k.a., slices) required to achieve FP64-level accuracy. Built on ESC, ADP integrates exception handling, run time heuristics, and seamless fallback to native FP64, ensuring correctness without host–device synchronization or user intervention. Additionally, we further improve Ozaki-style decompositions with an unsigned integer slicing scheme, which increases representational efficiency and reduces computational waste. Validated against recently proposed BLAS grading tests, ADP consistently preserves FP64 fidelity on challenging inputs while incurring less than 10\% run time overhead. In a 55-bit mantissa setting, our approach achieves up to 2.3$\times$ and 13.2$\times$ speedups over native FP64 GEMM on NVIDIA Blackwell GB200 and the RTX Pro 6000 Blackwell Server Edition, respectively. Our results demonstrate that low-precision accelerators can serve as a practical, production-ready foundation for high-fidelity and high-performance scientific computing workloads.
\end{abstract}





\received{20 February 2007}
\received[revised]{12 March 2009}
\received[accepted]{5 June 2009}

\maketitle

\section{Introduction}

Over the past decade, the rapid adoption of artificial intelligence has shifted GPU hardware design toward
\emph{low-precision formats} such as FP16, FP8, INT8, and, most recently, block-scaled MXFP8 and MXFP4 formats~\cite{ocp_mxfp23}. Specialized accelerators such as
Tensor Cores now deliver orders-of-magnitude higher matrix multiplication throughput in these formats than in traditional FP32 and FP64 pipelines~\cite{dongarra2024hardwaretrendsimpactingfloatingpoint,Kashi2024Survey}. This rift has motivated a new line of algorithm research: using low-precision units to emulate double-precision accuracy, most notably through \emph{Ozaki-style decompositions} that split FP64 values into low-precision slices, perform multiplications at high throughput, and reassemble the result~\cite{Mukunoki2020DGEMM,0otomo2024dgemm,uchino2025perf,Ozaki2025SchemeII}.

Recent work has demonstrated that such approaches can achieve FP64 accuracy
on GPUs~\cite{Ootomo2024DGEMM,Uchino2025INT8}. 
However, existing implementations require the user to specify the decomposition parameters, dictated by the numerical properties of the input data, required to maintain accuracy; as such, these are \emph{unsafe} and \emph{impractical} for production deployment~\cite{abdelfattah2025analysisfloatingpointmatrixmultiplication}. 
By \emph{unsafe}, we mean they cannot guarantee FP64-level accuracy under all inputs. By \emph{impractical}, we mean they lack mechanisms to ensure efficiency---either wasting compute on benign inputs or failing to perform competitively on difficult ones. Without safeguards for both accuracy and efficiency, Ozaki scheme-based emulation can only be used by a limited set of applications or for research, and cannot replace existing matrix multiplication implementations currently used in applications that rely on hardware-native FP64 capabilities.

\subsection*{Our Contributions}

This work addresses these aforementioned gaps with four contributions that together make FP64 emulation both safe and efficient on modern GPUs:

\begin{itemize}

  \item \textbf{Bit-efficient slicing.} We introduce an \emph{unsigned integer slicing
  representation} that eliminates redundant sign bits, increasing mantissa utilization and
  improving the accuracy of Ozaki-style decomposition on INT8 Tensor Cores.
  
  \item \textbf{Exponent Span Capacity (ESC).} We describe a hardware-agnostic estimator that
  based on input data, conservatively determines the number of bits needed to achieve FP64 accuracy.
  ESC eliminates reliance on fixed slice counts and provides the theoretical basis for adaptive precision.

  \item \textbf{Automatic Dynamic Precision (ADP).} We develop a GPU-resident workflow built on ESC
  that integrates exception handling, run time heuristics, and seamless fallback to native
  FP64. ADP guarantees accuracy by construction while ensuring performance does not fall
  below that of native FP64.

  \item \textbf{Structured validation.} We validate ADP with recently introduced
  \emph{BLAS grading tests}~\cite{Demmel2024,demmel2025aggressive}, providing a reproducible and rigorous assessment of numerical behavior across adversarial inputs.

\end{itemize}

The techniques described in this paper are general and can be 
extended to support a variety of numerical schemes for FP64 emulation. 
In this work, we focus on Ozaki Scheme I~\cite{Ootomo2024DGEMM} as a representative basis, 
as it is widely studied and provides a clear foundation for evaluating 
ESC and ADP. Extensions to other formulations, such as Ozaki Scheme II~\cite{Ozaki2025SchemeII}, 
are possible but are beyond the scope of this study.

To encourage reproducibility and adoption, we will release open-source 
implementations of ESC and our Ozaki-I-based DGEMM together with this paper.

\subsection*{Results and Impact}

On recent NVIDIA GPU architectures, ADP consistently achieves
\emph{componentwise accuracy indistinguishable from FP64 GEMM (i.e., DGEMM)} while introducing less than
10\% overhead. In favorable cases, our approach accelerates \texttt{cuBLAS}'s~\cite{nvidiacublasSite} DGEMM by 2.3$\times$ and 13.2$\times$ on Blackwell GB200 and RTX Pro 6000 Blackwell Server Edition GPUs, respectively. Together, ESC and ADP establish a safe, efficient, and
production-ready path for FP64 emulation on low-precision accelerators, bridging AI hardware
trends with the accuracy demands of high-performance scientific computing workloads.

\section{Background and Related Work}

High-performance scientific computing has long depended on IEEE double-precision (FP64) arithmetic for accuracy, stability, and reproducibility. However, motivated by AI applications such as Large Language Models (LLMs)~\cite{micikevicius2022fp8formatsdeeplearning,ocp_mxfp23}, modern GPUs have increasingly prioritized low-precision matrix multiplication and accumulation (MMA) execution units called Tensor Cores for types such as FP16, BF16, INT8, FP8, and, most recently, block-scaled FP4, which deliver orders-of-magnitude higher throughput than native FP64 pipelines. This architectural imbalance has motivated a growing body of work on emulating FP64 arithmetic with low-precision units. 

\subsection{FP64 Emulation on Low-Precision Units} \label{methodsketch}

A major milestone in this area was the adaptation of the \emph{Ozaki scheme}, originally proposed for reproducible summations~\cite{ozaki2012reproducible}, to GPUs. The scheme partitions FP64 numbers into multiple fixed-point \emph{mantissa slices}, computes products slice-by-slice, and recombines the results. Mukunoki et al.\ demonstrated that this approach could deliver FP64-accurate DGEMM on NVIDIA Tensor Cores~\cite{Mukunoki2020DGEMM}. Subsequent work showed that Ozaki-style decomposition could be generalized to higher-precision formats, including binary128 emulation~\cite{Uchino2025Ozaki}, and adapted to GPU integer matrix multiplication units~\cite{Ootomo2024DGEMM}. Uchino, Ozaki, and Imamura further optimized the scheme to improve efficiency, achieving superior performance--per--watt compared to native FP64 hardware~\cite{Uchino2025Ozaki}.

\subsection{Mixed-Precision Frameworks}

In parallel, mixed-precision solvers have been proposed as an alternative path to FP64 accuracy. Iterative refinement schemes use low-precision GEMM kernels as inner solvers while correcting residuals in FP64~\cite{haidar2018harnessing}. Domke et al.\ examined the practicality of such approaches in matrix engines for HPC workloads~\cite{Domke2021MatrixEngines}. More recently, techniques such as \emph{MixPert}~\cite{Lin2024MixPert} have demonstrated systematic ways to compose low-precision GEMM kernels into effective FP64 emulators. Surveys confirm that Tensor Core acceleration with precision refinement is becoming a mainstream strategy for sustaining FP64-level accuracy at a significantly lower energy cost~\cite{Kashi2024Survey}.

\subsection{Recent Advances and Open Challenges}

Several recent contributions have refined FP64 emulation techniques. Lightweight, cache-aware refinements have been explored in LE-GEMM~\cite{Zhang2025LEGEMM}, while modular arithmetic has been used to improve slice utilization in Ozaki Scheme II~\cite{Ozaki2025SchemeII}. Other directions include portable, hardware-agnostic estimators for precision selection~\cite{abdelfattah2025analysisfloatingpointmatrixmultiplication} and INT8-based GEMM emulation for energy-sensitive scenarios~\cite{uchino2025perf}, the acceleration of quantum chemistry through combinations of emulation and mixed precision on AI-centric GPUs~\cite{Dawson2024ReducingPrecisionQuantumChemistry}, and low-order orthogonal voxel finite elements with INT8 tensor cores for GPU-based elastic wave propagation analysis~\cite{Ichimura2025WavePropagation}.

Together, these works show that \emph{emulation is viable in principle}: 
FP64 accuracy can be reconstructed from low-precision units, and several 
techniques improve the efficiency of doing so.  However, \emph{critical gaps remain} for deployment in production HPC software as a default implementation. The three limitations in current approaches are:
\begin{itemize}
  \item \textbf{Representational inefficiency.} Signed-slice 
        encodings waste capacity by redundantly storing sign bits, 
        forcing higher slice counts than necessary and reducing 
        computational efficiency.
    \item \textbf{No universal safety guarantees.} Other methods 
        use a fixed slice count that the user has to input and cannot ensure FP64-level accuracy under adversarial inputs or extreme exponent ranges.
    \item \textbf{Lack of a run time framework.} Existing methods 
        do not provide mechanisms to detect numerical corner cases 
        (e.g., NaN/Inf propagation), predict performance, or 
        decide adaptively between emulation and native FP64. 
        Without such guardrails, emulation remains risky outside 
        of controlled benchmarks.
\end{itemize}

These open challenges explain why Ozaki-style emulation has not 
yet been adopted in production numerical libraries. Bridging this gap 
requires both a theoretical foundation for safe precision selection 
and a practical system that integrates safety checks, performance 
heuristics, and fallbacks into a GPU-resident workflow. This 
motivation underpins our contributions in Sections~\ref{encoding},~\ref{esc},  
and ~\ref{adp}.

\section{Unsigned Slice Encoding} \label{encoding}

When decomposing FP64 matrices into integer slices for emulation on INT8 Tensor Cores, a naïve approach is to store every slice as signed 8-bit integers (s8). This introduces an inefficiency: each sub-leading slice redundantly encodes a sign bit, even though the sign of the overall number is already captured in the most significant slice. As a result, each sub-leading slice contributes only 7 additional bits of effective mantissa precision, rather than the full 8 bits. To achieve a target precision equivalent to FP64's 53-bit mantissa, this approach requires eight slices (i.e. 64 bit), increasing both storage and compute costs.

In principle, the sign need only be stored in the leading slice, with subsequent slices encoded as unsigned 8-bit integers (u8).  To compute the leading signed slice, round-to-negative-infinity rounding is used, ensuring that the remainder is positive for representations using unsigned integer slices.  Under such a scheme, the FP64 mantissa could be represented in just seven slices rather than eight---a 22\% reduction in compute overhead. This optimization is feasible, since NVIDIA's integer Tensor Cores natively support mixed signed--unsigned arithmetic ($\text{s8} \times \text{u8}$). We prototyped this approach and confirmed its correctness.

However, our production implementation adopts an alternative method that leverages properties of two's complement arithmetic to achieve the same effect while avoiding the need for mixed signed--unsigned paths. The key insight is that we encode the full range of a u8 slice by redistributing values across multiple s8 slices:
\begin{itemize}
  \item If the original u8 slice lies in $[0,127]$, it is stored directly in s8.
  \item If the u8 slice lies in $[128,255]$, it is re-expressed as $256 - x$, where $x \in [0,127]$. The higher-order slice is incremented by 256, while the lower slice stores $-x$ in s8.
\end{itemize}

Two's complement ensures that this remapping preserves bitwise compatibility: the negative s8 value $-x$ has the same bit-string representation as the original u8 value. For example, a mantissa contribution of 
\[
123 \times 256 + 200 \ (\text{u8})
\]
can be equivalently represented as
\[
124 \times 256 - 56 \ (\text{s8}),
\]
where both $200 \ (\text{u8})$ and $-56 \ (\text{s8})$ share the bit pattern \texttt{0b11001000} (see Figure \ref{fig:unsigned-slice-general} for an illustration of this approach).

This unsigned slice encoding eliminates wasted sign bits, reduces slice count, and avoids specialized hardware paths, enabling a more efficient use of integer Tensor Cores for high-precision emulation.

\usetikzlibrary{positioning,arrows.meta}

\begin{figure}[ht]
\centering
\resizebox{\columnwidth}{!}{%
\begin{tikzpicture}[font=\ttfamily,>=Latex,thick]

\node[draw,fill=blue!10,minimum width=3.5cm,minimum height=1cm] (u8low) {u8 slice in [0,127]};
\node[below=0.2cm of u8low] (u8lowlbl) {bit pattern unchanged};

\node[right=5cm of u8low,draw,fill=green!10,minimum width=3.5cm,minimum height=1cm] (s8low) {s8 slice = same value};
\node[below=0.2cm of s8low] (s8lowlbl) {bit pattern unchanged};

\draw[->] (u8low.east) -- (s8low.west) node[midway,above]{direct};

\node[above=0.3cm of u8low] {\textbf{Case 1: direct copy}};

\node[below=2.5cm of u8low,draw,fill=blue!10,minimum width=3.5cm,minimum height=1cm] (u8high) {u8 slice in [128,255]};
\node[below=0.2cm of u8high] {e.g., 200 (0b11001000)};

\node[right=5cm of u8high,draw,fill=green!10,minimum width=3.5cm,minimum height=1cm] (s8high) {s8 slice = -x};
\node[below=0.2cm of s8high] {e.g., -56 (0b11001000)};

\node[above=0.8cm of s8high,draw,fill=orange!10,minimum width=3.5cm,minimum height=1cm] (carry) {+256 to higher slice};

\draw[->] (u8high.east) -- ++(2.2,0) node[midway,above]{remap} |- (s8high.west);
\draw[->] (u8high.east) -- ++(2.2,0) |- (carry.west);

\node[above=0.3cm of u8high] {\textbf{Case 2: remapping with carry}};

\end{tikzpicture}
}
\caption{Unsigned slice encoding using two's complement arithmetic. 
Case 1: u8 values in [0,127] map directly to s8 without modification. 
Case 2: u8 values in [128,255] are remapped as $256-x$ with a $+256$ carry to the higher slice, while storing $-x$ in s8. 
Bit patterns are preserved, e.g., 200 (u8) $\equiv$ -56 (s8) = \texttt{0b11001000}.}
\label{fig:unsigned-slice-general}
\end{figure}
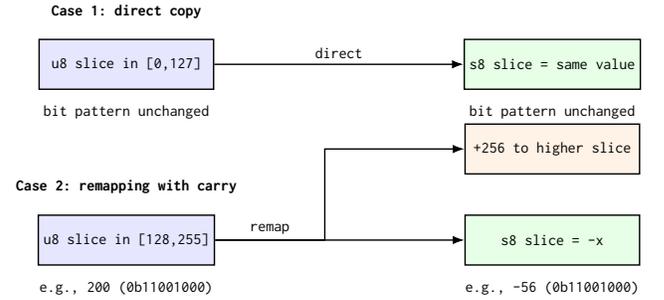

\section{ESC: Exponent Span Capacity Estimation} \label{esc}

When utilizing the Ozaki scheme, the number of bits needed to maintain the desired accuracy is dependent on the numerical contents of the constituent matrices~\cite{uchino2024performanceenhancementozakischeme}.  However, to provide users of emulated DGEMM with a familiar interface, it is necessary for the software to analyze the matrices and determine the number of bits needed in the intermediate representation.  We will refer to this as computing the Exponent Span Capacity (ESC).

As described in Section \ref{encoding}, our FP64 emulation is currently based on the use of integer-based (fixed-point) compute.  Because there is no exponent in this representation, when converting from FP64 to a multi-slice INT8 representation, a substitute for exponent functionality is required.  This involves shifting the mantissa bits in an extended storage format instead of altering an exponent field, shifting left for a greater exponent or right for one that is lesser.  For reasons of efficiency, the space to do this must be set aside at the outset of the matrix multiplication algorithm and must be uniform for the entire matrix. Assume, for example, that we are to preserve 53 mantissa bits during conversion and the greatest exponent field in a given row of the $A$ matrix is $00001100400$ (dec. 100).  For an entry in that row whose exponent is $100 - p$, we would set aside $p$ extra bits in the INT8 intermediate format.  The dynamic range of FP64 implies a potential storage requirement of approximately 2100 bits (the $2^{11}$ exponent bit range and the 53 additional bits needed to represent an IEEE FP64 mantissa with the implicit bit made explicit).  However, using such a fixed representation would entail a prohibitive amount of both storage and emulative compute, as the Ozaki-I scheme's compute requirements are quadratic in the number of slices used.  However, it is necessary to determine how many extra "padding bits" are needed to compute with the same precision used by IEEE FP64, without using excess resources.

The formula used to compute the ESC is deterministic, reproducible, and easy to understand, both intuitively and formally.  Since matrix multiplication, $C_{m \times n} = A_{m \times k} B_{k \times n}$, is an unordered set of dot products, we will isolate and examine the ESC for a single dot product, $x^{T} \cdot y$, where $x^{T}$ is a row vector of $k$ FP64 values and $y$ is a column vector of $k$ FP64 values.

The ESC for matrix multiplication is the maximum ESC of the $mn$ component dot products.  For purposes of exposition, it is helpful to view the dot product evaluation as a two-step process, though these operations are fused in practice.  These steps are:

\begin{enumerate}
\item Multiply each $x_i$ $\odot$ $y_i$ $\rightarrow$ $z_i$, the Hadamard product
\item Reduce all $z_i$ $\rightarrow$ $s$, a scalar value ($s = \sum_{i=1}^k z_i$) 
\end{enumerate}
In computing the ESC, the mantissa field is ignored, so all values with the same exponent are viewed as equivalent and the elements of $z$ are the result of adding the exponents of the elements in $x$ and $y$, not element-wise multiplication.  We will discuss why this is safe to do after the formula is presented.  There are five designated values of interest:
\begin{enumerate}
\item $x_p$, entry with the maximum exponent in $x$ 
\item $y_q$, entry with the maximum exponent in $y$ 
\item $z_r$, entry with the maximum exponent in $z$ 
\item $x_r$, entry in $x$, giving rise to $z_r$ 
\item $y_r$, entry in $y$, giving rise to $z_r$
\end{enumerate}

Note that $p$, $q$, and $r$ may not be unique, but the exponents of $x_p$, $y_q$, and $z_r$ are each sets with a single value.  

The ESC is calculated as
ESC $=$ exp($x_p$) $+$ exp($y_q$) $-$ exp($z_r$), where exp($z_r$) $=$ exp($x_r$) $+$ exp($y_r$) and exp($\cdot$) denotes the exponent component of $\cdot$.

The algorithm described above is for a single dot product in a matrix multiplication composed of $mn$ dot products of length $k$.  For performance reasons, it is highly desirable to reduce the time to compute the ESC without sacrificing the safety it provides.  This is done by computing an estimate of $z_r$ as follows.  Both the $x$ and $y$ vectors are coarsened, broken into blocks of length $b$, where the greatest and least exponent in each block, Max($b_i$) and Min($b_i$) are stored as representatives of that block.  Then, $z_r$ is approximated by finding the maximal value, over all blocks, $i$, of Max($xb_i$)+Min($yb_i$) and Min($xb_i$)+Max($yb_i$).  This may underestimate $z_r$, resulting in a larger ESC value than the non-coarsened, but computationally impractical, version of the algorithm.  However, the primary goal is safety and it is straightforward to demonstrate that the coarsened algorithm cannot overestimate the exponent of $z_r$, derived from the non-coarsened algorithm, as follows. 

Assume, for some block index, $i$, $z_r < $ Max($xb_i$) +  Min($yb_i$), as the other case is symmetric.
Let exp($x_j$) $ = $ Max($xb_i$) for one or more $j$ in the span of $xb_i$.
For all such $j$, 
exp($x_j$) $+$ Min($yb_i$) is greater than exp($x_j$) $+$ exp($y_j$) by the assumption that the non-coarsened ESC, $z_r$, is less than the coarsened ESC.  Also, Min($yb_i$) $\leq$ exp($y_j$) by definition.  Therefore, exp($x_j$) $+$ Min($yb_i$) $\leq$ exp($x_j$) $+$ exp($y_j$) and exp($x_j$) $+$ exp($y_j$) $\leq z_r$ by the definition of $z_r$.  Given that exp($x_j$) $=$ Max($xb_i$), by transitivity,  Max($xb_i$) $+$ Min($yb_i$) $\leq z_r$, contradicting the assumption.

This formulation supplies us with some valuable guardrails.  First, it gives us a guarantee that the maximal (in terms of exponent magnitude) contribution to the dot product is captured with full fidelity.  That is, the number of bits dictated by the user (the default being 53 bits), are all used in the calculation of $z_r$.  Put another way, all of the mantissa bits of $x_r$ and $y_r$ will be extracted from the corresponding FP64 input values, placed in fixed-point components, and used to calculate $z_r$ using the Ozaki scheme.  This is because we will reserve enough fixed-point slices to accommodate the requisite mantissa bits as well as all of the placeholder bits needed for shifting these representations, as described above.   Notice that this applies to all such contributions to the dot product.  As $r$ is not necessarily unique, exp($z_r$) $=$ $F$, can arise from multiple exp($x_r$) and exp($y_r$) values.  This formula gives the “full fidelity guarantee” for all combinations of these exponents, where each summand is the same (maximal) value, $F$, and simultaneously removes any need to track the indices, $r$.  Since the resultant emulation utilizes all source mantissa bits for even the most asymmetric pairings, it also captures the “funnel” of values around $F$ (i.e. the method will discard no more than $j$ bits when the Hadamard exponent is $F-j$).  This is also the reason why the mantissa field can be ignored, as regards $z_r$.  Multiplying the mantissa values of two scalars can result in an exponent increase of 1 (the product of the mantissas is always less than $4.0$), we increment the ESC value by $1$ to guarantee the margin of safety.

Assuming that all values are of the same sign, this approach yields the same precision characteristics as IEEE FP64.  However, the cancellation characteristics are different.  Emulation results are invariant with respect to the ordering of summed values, while those utilizing IEEE FP64 arithmetic are not.  The strength of native FP64 in this regard rests in the dynamic exponent.  There are instances where the shifting of the exponent, due to either gradual or dramatic cancellation, provides an advantage.  In other cases, the extended precision enabled by the extra bits utilized in the intermediate form of the summand will lead to improved accuracy.


\section{ADP: Automatic Dynamic Precision}  \label{adp}

While emulation provides a path to recover FP64 accuracy on low-precision Tensor Cores, its practicality depends on integrating precision estimation, numerical safety checks, and performance-aware run time decisions. In this section we describe our fully GPU-resident workflow, which leverages the Exponent Span Capacity (ESC) estimator together with lightweight pre-processing kernels to enable seamless run time integration.

\subsection{Pre-processing and Safety Checks}

The implementation described in this paper adheres to FP64 levels of accuracy for both normal and denormalized values, but does not strictly adhere to the IEEE 754 standard in its treatment of signed zeros, \texttt{Inf}s, and \texttt{NaN}s.  Negative zeros in the input are simply treated as a zero, since our implementation does not handle signed zeros. Our handling of emergent \texttt{NaN}s and \texttt{Inf}s is similar to IEEE 754 and is described below. However, a key limitation of Ozaki-style emulation is its inability to handle input that includes special values such as \texttt{Inf} and \texttt{NaN}. If left unchecked, these values propagate incorrectly through slice decompositions, leading to undefined behavior. To ensure numerical safety, we scan matrices $A$ and $B$ prior to multiplication. If either scan detects \texttt{Inf} or \texttt{NaN}, the workflow falls back to native FP64 GEMM, guaranteeing correctness. In practice, this scanning occurs while preparing for the coarsened ESC calculation and fallback can be incurred before any $O(n^3)$ algorithms are executed.  The separate treatment of the exponent in the Ozaki method leads to emergent signed \texttt{Inf}s that reflect the \texttt{Inf} of greater magnitude and does not produce emergent \texttt{NaN}s in all cases where FP64 would do so.
  While it is possible to anticipate (imperfectly) the possibility of emergent \texttt{NaN}s and \texttt{Inf}s, possibly invoking FP64 fallback in such cases, we do not do this.  The Ozaki algorithm can give rise to +Inf and \texttt{-Inf} when the result of the dot product is converted back to an FP64 representation, but \texttt{Inf}s are not ``sticky,'' as this is the terminal step in our emulation algorithm.  The benefit to this is that accumulated values may be brought back into the normal range of FP64 values.  Though the Ozaki method itself does not address emergent \texttt{NaN}s, the necessity to aggregate partial results, so as to avoid overflowing accumulators, can give rise to sticky \texttt{Inf}s and \texttt{NaN}s if that intermediate aggregation includes conversion to FP64 values.  Our implementation will generate \texttt{Inf}s when they are necessary, though not in all of the cases where an FP64-based computation would do so, and \texttt{NaN}s in a subset of the instances where FP64 would generate \texttt{NaN}s, given the disparate generation of emergent \texttt{Inf}s.

\subsection{Exponent Span Estimation}

The exponent span estimator begins by pre-processing the $A$ and $B$ matrices to find the max and min exponent per block for our coarsened Exponent Span Capacity.  After the inputs have been preprocessed, a dedicated kernel implements the coarsened exponent span capacity.  The keen observer may notice that this $O(n^3)$  algorithm and is reminiscent of a GEMM.  We extend the \texttt{CUTLASS} library~\cite{cutlass} and leverage dynamic programming (DPX) instructions~\cite{dpx} available on Hopper and newer GPUs to significantly accelerate this process.  Thanks to these hardware features and safety guarantees, we are able to have an accurate \emph{and} cost-effective method to estimate the exponent span capacity.

\subsection{Heuristic run time Selection}

After ESC reports the required bit width, a lightweight heuristic evaluates whether emulation is expected to outperform native FP64. If the slice count remains within a performance-efficient range, the workflow dispatches the emulated GEMM kernel. Otherwise, native FP64 GEMM is launched instead. This mechanism ensures that emulation is only applied when beneficial, avoiding performance regressions for difficult cases with large exponent spans.

\subsection{GPU-Resident Integration}

Crucially, all of the above steps—safety scans, ESC estimation, and heuristic selection—are implemented as CUDA kernels running entirely on the GPU. No host--device synchronization is required, allowing our approach to integrate transparently with existing GPU libraries such as \texttt{cuBLAS}. The result is a run time framework that provides FP64-accurate GEMM with dynamic precision tuning and negligible decision overhead, while ensuring correctness in the presence of exceptional values.


\section{Numerical Accuracy}

This section evaluates the numerical accuracy of our implementation of ADP-enabled DGEMM in the \texttt{cuBLAS} library. 
We demonstrate that our implementation with guardrails enabled indistinguishably behaves like a native floating-point implementation.

It is well understood that different implementations of matrix multiplication satisfy different accuracy bounds.
In response to the appearance of fixed point implementations~\cite{0otomo2024dgemm}, Demmel et al.~\cite{Demmel2024,demmel2025aggressive} devised tests that detect the type of the underlying matrix multiplication algorithm. 

They considered two dimensions. First, an implementation can either be $\mathcal{O}(n^3)$ or Strassen-like.
Second, it can use either floating-point or fixed point arithmetic. 
The following tests, when executed as a tree, cover the four possible combinations.

\begin{itemize}
\item \textbf{Test~1}: Distinguish a conventional \begin{math}\mathcal{O}(n^3)\end{math} implementation from a Strassen-like implementation.
\item If Test~1 detects an \begin{math}\mathcal{O}(n^3)\end{math} implementation, use \textbf{Test~2}: Distinguish a \begin{math}\mathcal{O}(n^3)\end{math} floating-point implementation from an \begin{math}\mathcal{O}(n^3)\end{math} fixed point implementation.
\item If Test~1 detects a Strassen-like implementation, use \textbf{Test~3}: Distinguish a Strassen-like floating-point implementation from a Strassen-like fixed point implementation. 
\end{itemize}
The tests rely on differences in the computed results to determine the type of algorithm, and not on (asymptotic) run time measurements. 

In addition to the algorithm discovery, they summarized accuracy criteria for further assigning a grade to a matrix multiplication implementation. 
We evaluate the grade~A criterion, which measures the componentwise relative error.

Following the work by Demmel et al.~\cite{Demmel2024}, we demonstrate the following two aspects through numerical experiments.
\begin{enumerate}
\item[A1] Test~2 cannot distinguish the Ozaki based DGEMM from an $\mathcal{O}(n^3)$ floating-point implementation if it has the guardrails enabled and may fall back to native FP64 DGEMM. This is the default behavior.
\item[A2] Ozaki based DGEMM with automatic fallback meets the grade A criterion and attains an accuracy similar to native $\mathcal{O}(n^3)$ FP64 DGEMM implementations provided in highly optimized libraries.
\end{enumerate}

\paragraph{Aspect A1} 
Test~2 is most important to our evaluation since the Ozaki based DGEMM is an $\mathcal{O}(n^3)$ implementation (and not Strassen-like).
Test~2~\cite{demmel2025aggressive} exploits the fact that a fixed point implementation can lose accuracy when not all necessary bits are covered.
For that reason, the test chooses the input matrices to have a wide exponent span.
As a result, fixed point implementations that use a predetermined, fixed number of bits likely do not cover all bits required for an accurate computation.

Specifically, the matrices \begin{math}\boldsymbol{A} \in \mathbb{R}^{n \times n}\end{math} and \begin{math}\boldsymbol{B} \in \mathbb{R}^{n \times n}\end{math} are constructed starting from a vector with entries uniformly distributed in (1,2)
\begin{align*}
\boldsymbol{x} \sim \mathcal{U}(1,2)^{n \times 1}.
\end{align*}
The exponent range is determined by the diagonal matrix
\begin{align*}
\boldsymbol{D} = \operatorname{diag}(2^{j_1}, \hdots, 2^{j_n}),
\end{align*}
where the exponents of the diagonal entries are chosen as integers
\begin{align*}
j_{i+1} = -b + \mathrm{round}(i\Delta).
\end{align*}
Demmel et al.~\cite{demmel2025aggressive} select \begin{math} b \end{math} and \begin{math} \Delta = 2b/(n-1) \end{math} such that \begin{math} -j_1 = j_n \end{math} and \begin{math} b \approx \lfloor\log_2(\sqrt{\Omega})\rfloor-\lceil\log_2(n)\rceil-1 \end{math}, where \begin{math}\Omega \end{math} is the overflow threshold.
We generalize Test~2 and treat \begin{math}b \end{math} as a parameter that controls the exponent range.

Then, the \begin{math}k\end{math}th row of \begin{math}\boldsymbol{A}\end{math} and the \begin{math}k\end{math}th column of \begin{math}\boldsymbol{B}\end{math} are computed as
\begin{align*}
\boldsymbol{A}_{k,:} = \boldsymbol{x}^T \boldsymbol{D} \boldsymbol{P}_k, \quad \boldsymbol{B}_{:,k} = \boldsymbol{P}^{-1}_k \boldsymbol{D}^{-1} \boldsymbol{x}.
\end{align*}
\begin{math}\boldsymbol{P}_k\end{math} is the permutation matrix that realizes a cyclic row shift of \begin{math}\boldsymbol{A}\end{math} by \begin{math}k \end{math} entries. The permutations ensure that implementations cannot game the test by strategically rescaling the input matrices.

\begin{figure}

\includegraphics[width=\columnwidth]{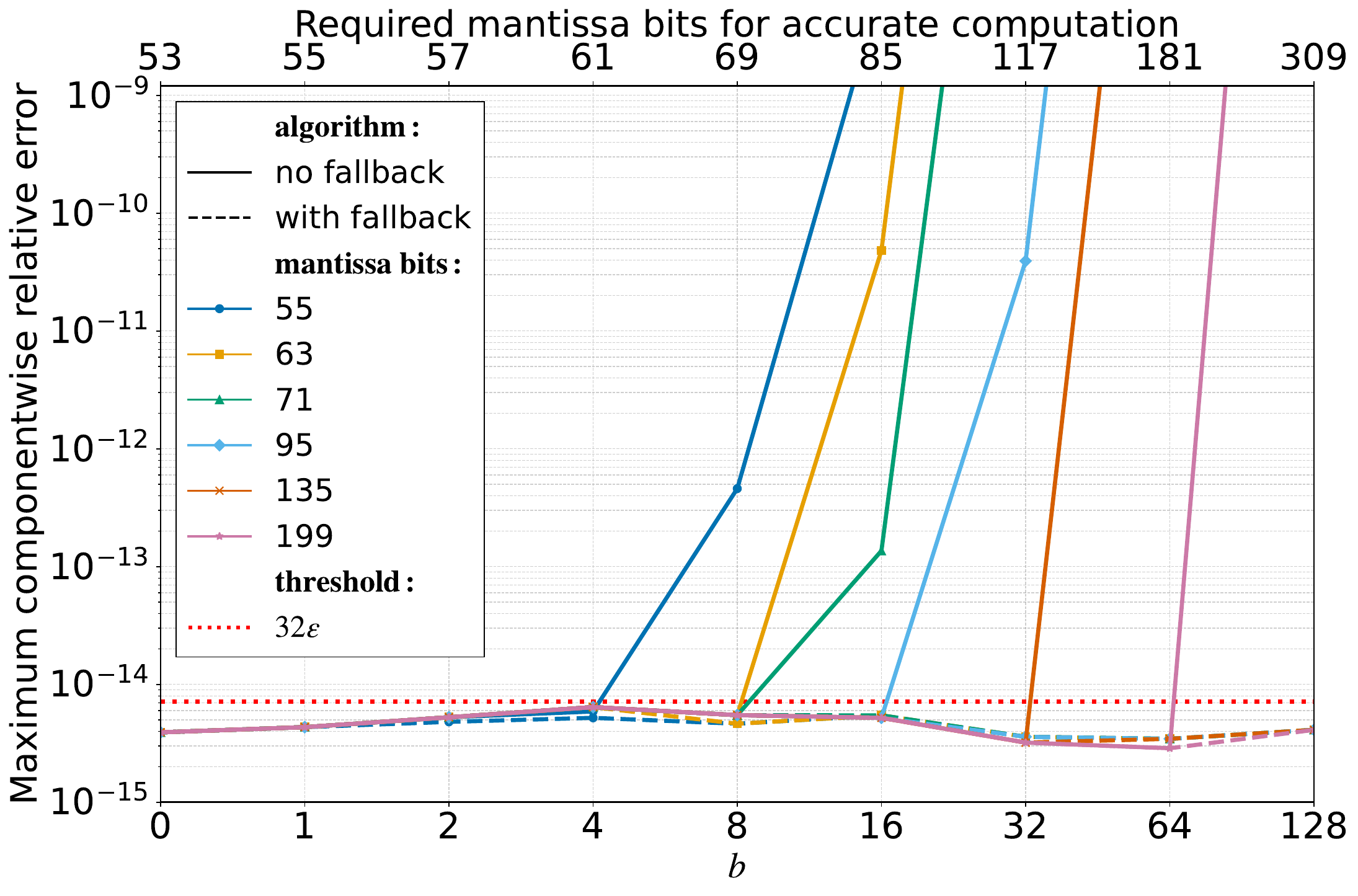}

\caption{ADP-enabled DGEMM, configured with six distinct mantissa bit counts for the Ozaki-I algorithm, on Test 2, where \begin{math}n = 1024\end{math}.
For each mantissa bit count, we display the variant without the option to fall back to native FP64 DGEMM (solid lines) and the variant with guardrails and automatic fallback to native FP64 DGEMM (dashed lines).}
\label{fig:accuracy-test2}
\end{figure}

\autoref{fig:accuracy-test2} shows the results of ADP-enabled DGEMM with and without guardrails and automatic fallback, respectively, on Test~2 for increasing values of \begin{math}b \end{math}.
The relative error is computed as
\begin{align*}
\max_{ij} e_{ij}, \quad
e_{ij} =
\begin{cases}
\frac{|\boldsymbol{x}^T\boldsymbol{x} - c_{ii}|}{|\boldsymbol{x}^T\boldsymbol{x}|}, & \text{if }i = j\\
\frac{|c^\text{ref}_{ij} - c_{ij}|}{|c^\text{ref}_{ij}|}, & \text{otherwise.}
\end{cases}
\end{align*}
For that, we use a reference \begin{math}\mathcal{O}(n^3)\end{math} floating-point implementation of GEMM to compute \begin{math}\boldsymbol{C}^\text{ref} = \boldsymbol{A}\boldsymbol{B}\end{math}.
In addition, we compute the diagonal entries through \begin{math}\boldsymbol{x}^T\boldsymbol{x}\end{math} using long double (FP80).

Our implementation of emulated DGEMM based on Ozaki-I scheme, by default, falls back to native FP64 DGEMM if emulation is infeasible.
For Test~2, this occurs when the ESC exceeds the available number of bits.
For this type of matrix, the ESC with and without coarsening computes the required number of mantissa bits accurately without overestimation.
Without guardrails, the emulated DGEMM reaches a high relative error if \begin{math}b\end{math} is sufficiently large and the available mantissa bits cannot represent all bits required for an accurate computation and, hence, fails Test~2.
As such, Test~2 as designed by Demmel et a~\cite{Demmel2024,demmel2025aggressive} is reliable.
With guardrails and fallback option, the emulated DGEMM attains a small relative error comparable to what can be expected from as a standard \begin{math}\mathcal{O}(n^3)\end{math} floating-point GEMM.

\paragraph{Aspect A2}
Implementations of GEMM can be evaluated according to different accuracy criteria.
Demmel et al.~\cite{Demmel2024} established a framework that defines three grades.
Grade~C is the weakest grade and is defined as a norm-wise bound that can be satisfied by Strassen-like algorithms.
Grade~B is a mixed componentwise and norm-wise bound, which is invariant under certain diagonal scalings.
Grade~A represents the most stringent criterion.
It considers the componentwise relative error bound
\begin{align}\label{eq:gradeA}
| \operatorname{fl} (\boldsymbol{A}\boldsymbol{B})_{ij} - (\boldsymbol{A}\boldsymbol{B})_{ij} | \leq f(n) \epsilon (|\boldsymbol{A}| |\boldsymbol{B}|)_{ij},
\end{align}
where \begin{math}f(n)\end{math} measures how the error accumulates.
Inequality~\eqref{eq:gradeA} is considered the gold standard for the evaluation of numerical accuracy and is used by Higham~\cite[Eq. (3.13)]{higham2002accuracy} and reference (``netlib'') \texttt{LAPACK}~\cite{10.5555/323215}.

The function \begin{math}f(n)\end{math} in inequality~\eqref{eq:gradeA} is critical for the assessment of how accurate a GEMM implementation is.
For grade~A compliance, \begin{math}f(n)\end{math} must not exceed linear growth.
Miller~\cite{miller1974computational} demonstrates that enforcing strong numerical stability in matrix multiplication -- defined as bounded error growth relative to input size -- necessitates $O(n^3)$ computational complexity, thereby excluding faster algorithm such as Strassen's from satisfying such stability guarantees.

In our numerical experiments, we compare the accuracy of emulated DGEMM, native FP64 DGEMM, and a floating-point Strassen matrix multiplication. 
Emulated DGEMM is configured to use up to 200 mantissa bits.
We have validated that the emulated DGEMM never falls back to native FP64 DGEMM in any of the experiments.
The floating-point Strassen implementation is a simple reference implementation that we include for comparison purposes.
The experiment multiplies two square matrices whose entries are uniformly distributed in (0,1).
Each routine is executed with five distinct seeds.
\begin{figure}
\centering
\includegraphics[width=\columnwidth]{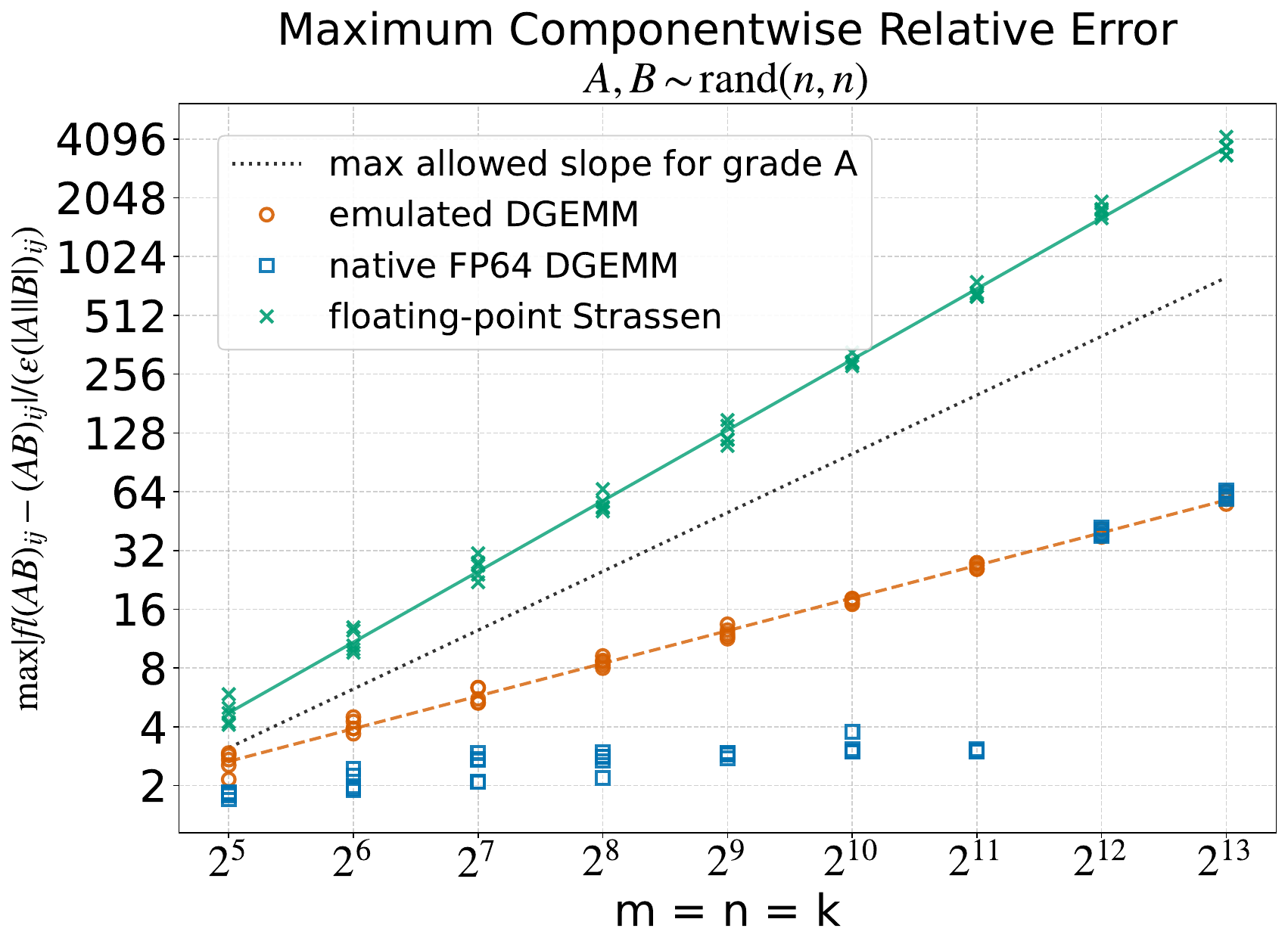}
\Description{Analysis of the maximum relative error on two random uniformly distributed matrices.}
\caption{Maximum componentwise relative error when multiplying two random uniformly distributed matrices.}
\label{fig:accuracy-maxerr}
\end{figure}
Figure~\ref{fig:accuracy-maxerr} shows the maximum componentwise relative error.
Emulated DGEMM consistently meets the Grade A criterion across all matrix sizes, with the maximum componentwise relative error growing well below the maximum allowed slope.
The floating-point Strassen implementation, as expected, exceeds the maximum slope allowed for Grade A.
Native FP64 DGEMM changes its error accumulation behavior at $n = 4096$, which indicates an algorithm switch.
For large matrices, the emulated DGEMM and native FP64 DGEMM show comparable error accumulation behavior.

\begin{figure}
\centering
\includegraphics[width=\columnwidth]{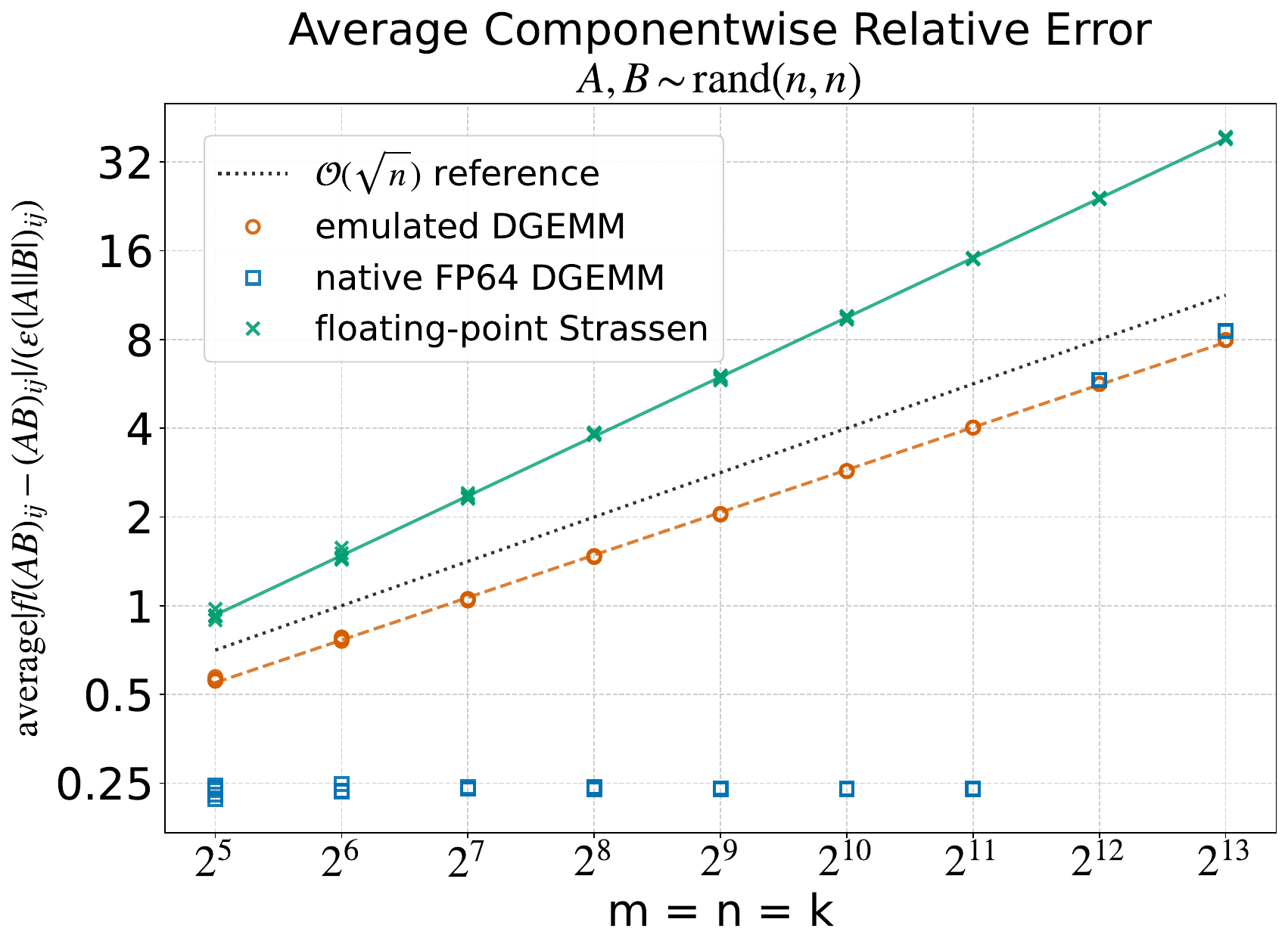}
\Description{Analysis of the average relative error on two random uniformly distributed matrices.}
\caption{Average componentwise relative error when multiplying two random uniformly distributed matrices.}
\label{fig:accuracy-avgerr}
\end{figure}
Figure~\ref{fig:accuracy-avgerr} shows the average componentwise relative error.
The error pattern of emulated DGEMM closely follows the theoretically expected growth proportional to $\sqrt{n}$ for random uniformly distributed matrices.
For larger matrices, native FP64 DGEMM follows the same expected pattern.
This experiment supports the claim that our implementation of Ozaki DGEMM accumulates errors in a manner similar to native double-precision arithmetic and as dictated by theoretical error analysis.

\section{Performance Evaluation}
\label{sec:performance}

We evaluate ADP-enabled DGEMM on recent NVIDIA architectures 
to assess both its efficiency and deployability. The experiments 
cover microbenchmarks, end-to-end DGEMM throughput, and 
integration into NVIDIA's \texttt{cuSOLVER}~\cite{nvidiacusolverSite} library.
The results highlight that ADP’s 
run time guardrails incur modest overhead while preserving most 
of the performance gains of emulation, and that these benefits 
extend transparently to HPC workloads.

\subsection{Performance Breakdown}
\label{sec:breakdown}

\begin{figure}
\centering
\includegraphics[width=\columnwidth]{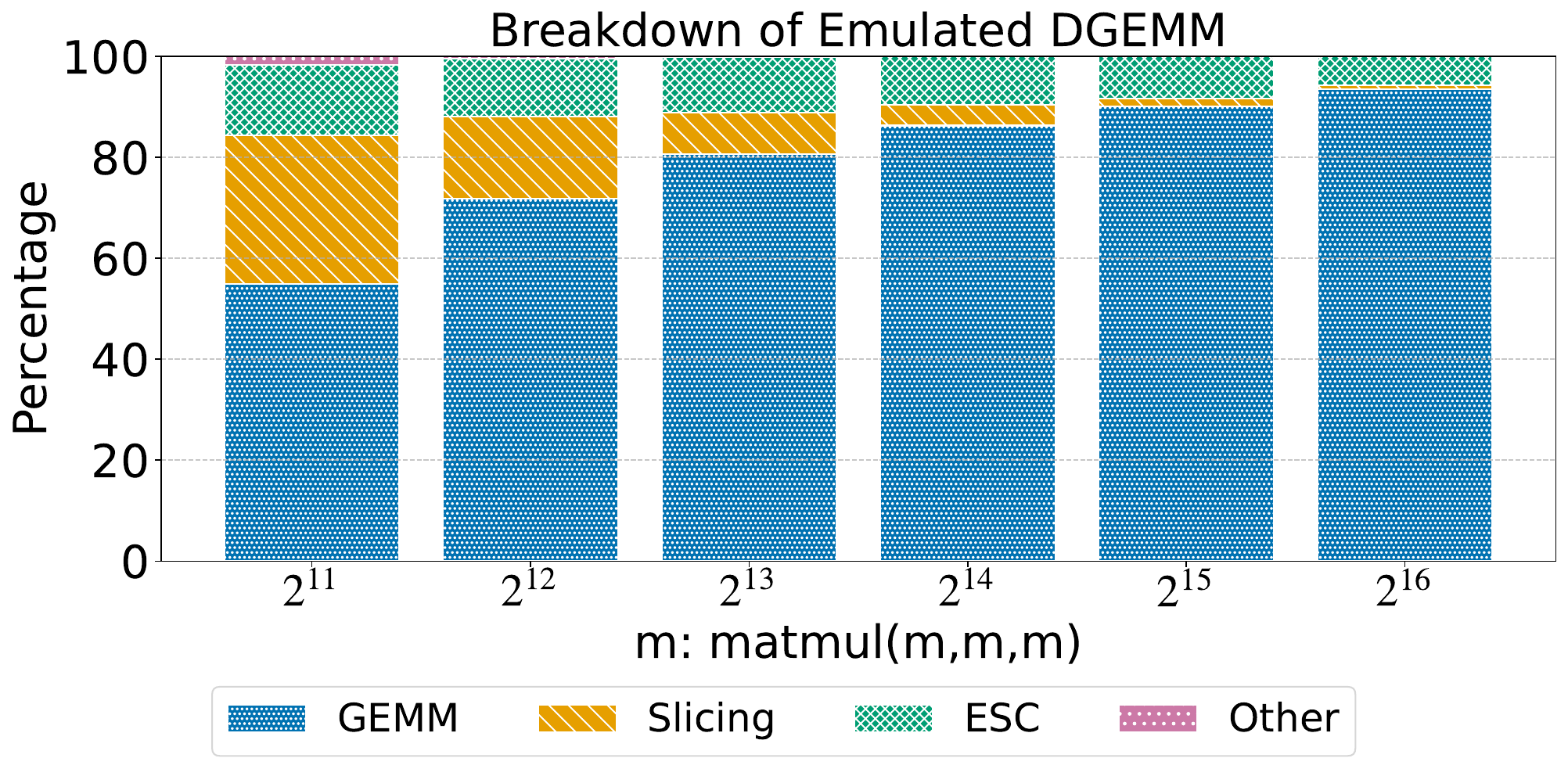}
\Description{Breakdown of DGEMM wall time.}
\caption{Breakdown of DGEMM performance when emulating 55 mantissa bits 
on NVIDIA Tensor Cores. For this experiment, ADP is forced to always use 
55 bits, regardless of input characteristics, in order to maximize its 
relative overhead (worst-case configuration).}
\label{fig:breakdown-esc-NN}
\end{figure}

Figure~\ref{fig:breakdown-esc-NN} reports the breakdown of DGEMM performance 
when ADP is configured to emulate 55 mantissa bits. In this experiment, ADP 
is forced to always use 55 bits, regardless of input characteristics, in order 
to expose its overhead under the most challenging conditions.

This setup deliberately places ADP at its largest disadvantage. The cost of 
ADP’s guardrails (fallback checks, exception handling, heuristics) is essentially 
constant, while the cost of slicing and low-precision multiplications grows with 
the number of mantissa bits. By fixing the precision at 55 bits, the relative 
share of ADP overhead is maximized. For smaller mantissa counts, the total 
computation shrinks but ADP’s cost remains flat, so its relative contribution 
is higher.

This experiment confirms that even in this worst-case configuration, ADP 
adds less than 10\% overhead to the total DGEMM run time. The majority of 
execution time is spent in integer matrix multiplications and slice 
recomposition, while ADP’s run time checks account for only a small and 
predictable fraction of cost.

Overall, Figure~\ref{fig:breakdown-esc-NN} demonstrates that ADP overhead is modest and bounded: 
even when forced to operate at 55 mantissa bits, where its relative impact is 
maximized, ADP consumes less than 10\% of run time. In more typical cases, its 
cost is smaller still, making ADP practical as a default safeguard in FP64 
emulation.

\subsection{Performance}
\label{sec:endtoend}

\begin{figure*}[htbp]
  \centering
  \begin{subfigure}{\textwidth}
    \includegraphics[width=\linewidth]{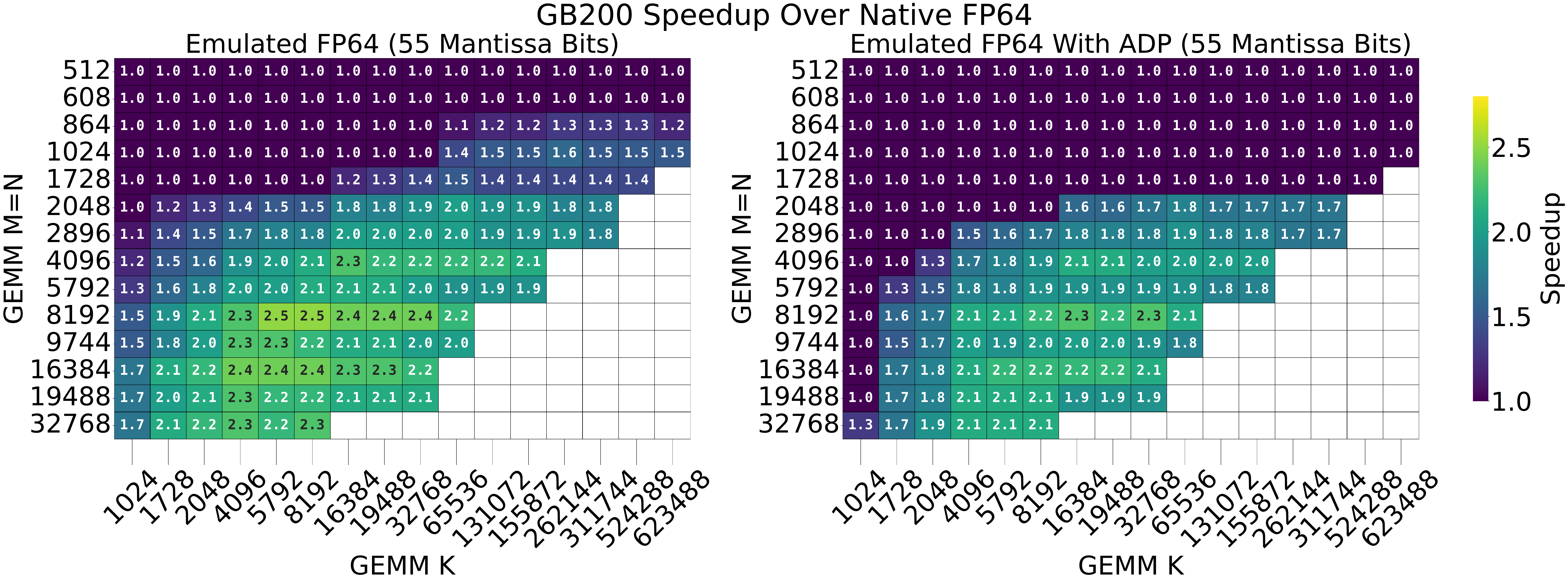}
  \end{subfigure}
  
   \begin{subfigure}{\textwidth}
    \includegraphics[width=\linewidth]{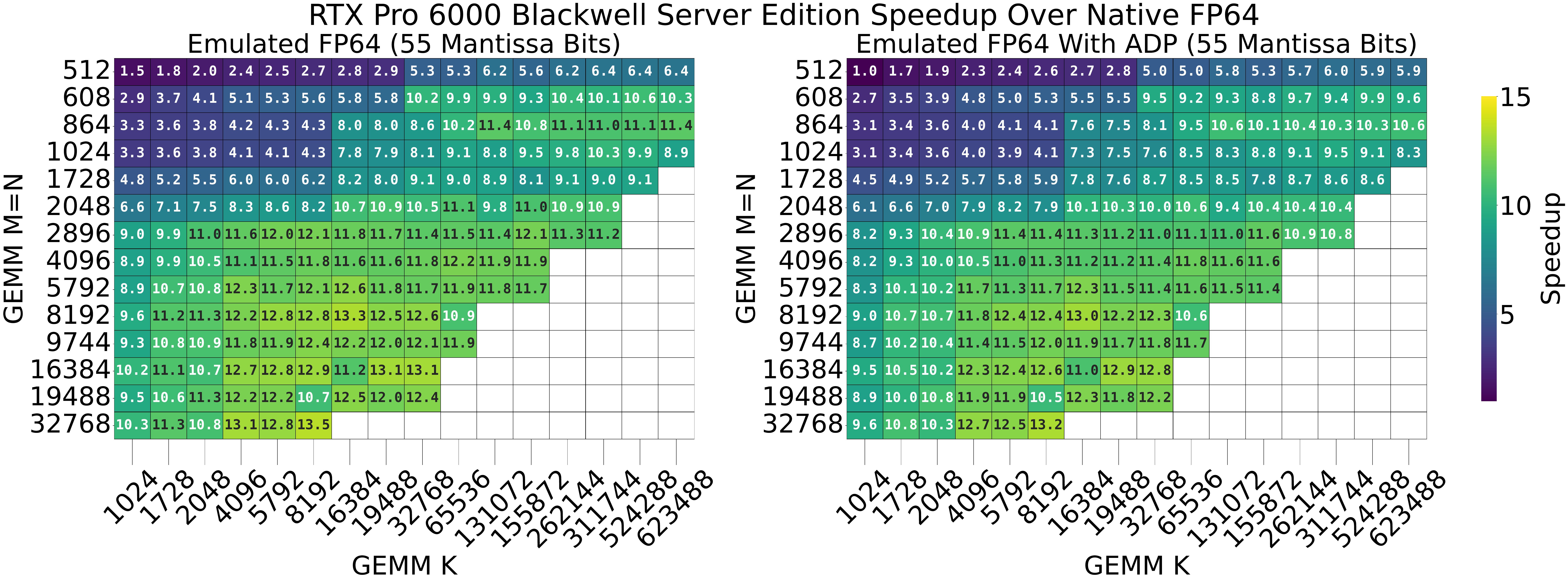}
  \end{subfigure}

  \caption{End-to-end DGEMM performance comparison on two GPU platforms: 
(top) GB200 and (bottom) RTX Pro 6000 Blackwell Server Edition. 
Each row shows speedups over cuBLAS's native DGEMM for two cases: 
(left) emulated DGEMM with 55 mantissa bits and no ADP (upper bound on performance, but no safety guarantees)
and (right) emulated DGEMM with ADP forced to use 55 mantissa bits.}
  \label{fig:cublasdgemm_endtoend}
\end{figure*}

Figure~\ref{fig:cublasdgemm_endtoend} summarizes end-to-end GEMM speedups over
cuBLAS's native DGEMM
for two GPU NVIDIA platforms: GB200 (top row) and RTX Pro 6000 Blackwell Server Edition 
(bottom row). Each row presents two values: (left) emulated DGEMM without ADP and (right)
emulated DGEMM with ADP forced to use 
55 mantissa bits. 

The left plot establishes the performance upper bound of this experiment: emulated DGEMM 
with 55 mantissa bits and no guardrails achieves the highest throughput, but 
does not provide any safety guarantees.
The right chart shows the same 
emulation with ADP enabled. Here, performance is nearly identical to the 
no-ADP baseline, with only a small difference accounting for run time guardrails. 
This confirms that ADP preserves almost all of the performance benefit of 
emulation while adding accuracy guarantees.

The bottom row shows analogous results on the RTX Pro 6000 Blackwell 
Server Edition. Again, the left plot (55-bit emulation without ADP) 
represents the performance ceiling, and the right chart shows that ADP 
closely matches this ceiling while ensuring safe deployment. Across both 
architectures, the additional cost of ADP remains well below 10\% of 
total run time, even under the forced 55-bit configuration where its 
overhead is maximized.

Taken together, Figure~\ref{fig:cublasdgemm_endtoend} demonstrates that ADP is practical and 
deployable: it enables emulated DGEMM to outperform native FP64 by 
up to 2.3$\times$ and 13.2$\times$ speedups over
native FP64 GEMM on NVIDIA Blackwell GB200 and the RTX Pro 6000 Blackwell Server Edition, respectively, while guaranteeing accuracy.

\subsection{Application-Level Integration}
\label{sec:application}

To evaluate the practicality of ADP beyond synthetic DGEMM
benchmarks, we integrated our framework into NVIDIA’s \texttt{cuSOLVER} library, and specifically the 
\texttt{cusolverDnGeqrf} routine. The QR factorization in \texttt{cuSOLVER}
is implemented using the compact $WY$ representation of block 
Householder reflectors~\cite{schreiber1989wy}, as outlined in 
Algorithm~\ref{alg:qr_algorithm}. Lines \ref{line:QR_update_begin}–\ref{line:QR_update_end} correspond to the trailing 
matrix update---the BLAS3 component that dominates run time 
for medium and large problem sizes---which we redirected to use 
emulated DGEMM with ADP guardrails.

\begin{algorithm}
    \begin{flushleft}
    \textbf{Input:} $A\in\mathbb{R}^{m\times n}$, assume $\min(m,n)$ is an integer multiple of the panel width $b$\\
    \textbf{Outputs:} Matrix of Householder reflectors $Y\in\mathbb{R}^{m\times n}$, upper triangular matrix $R\in \mathbb{R}^{m\times n}$, both overwriting $A$
    \end{flushleft}
    \begin{algorithmic}[1]    
        \State Partition $A \rightarrow \begin{bmatrix} A_{1} & A_{s} \end{bmatrix}$ where $A_{1}$ is $m \times b$
        \For{$i \gets 1, \ldots, \min(m,n)/b$}
            \State Factor $A_{i} = Q_i R_i$ such that $Q_i = I - Y_iT_iY_i^T$
            \State Store $A_i \gets \begin{bmatrix} \quad R_i  \\ \diagdown \\ Y_i \quad  \end{bmatrix}$       
            \State Update $A_s$ by performing $Q_i^TA_s$ as follows:
                \State \hspace{\algorithmicindent}$W \leftarrow Y_{i}^{T} A_s$ \Comment{GEMM} \label{line:QR_update_begin}
                \State \hspace{\algorithmicindent}$W \leftarrow T_i^{T} W$ \Comment{GEMM for efficiency}
                \State \hspace{\algorithmicindent}$A_s \leftarrow A_s - Y_{i} W$ \Comment{GEMM} \label{line:QR_update_end}
            \State Partition $A_s \rightarrow \begin{bmatrix} R_{i,2} & R_{i,3} \\ A_{i+1} & A_{s}\end{bmatrix}$ where $A_{i+1}$ is $m-ib \times b$
        \EndFor
    \end{algorithmic}
    \caption{Blocked Householder-QR factorization}
    \label{alg:qr_algorithm}
\end{algorithm}

\begin{figure*}[htbp]
\centering
\includegraphics[width=1.0\textwidth]{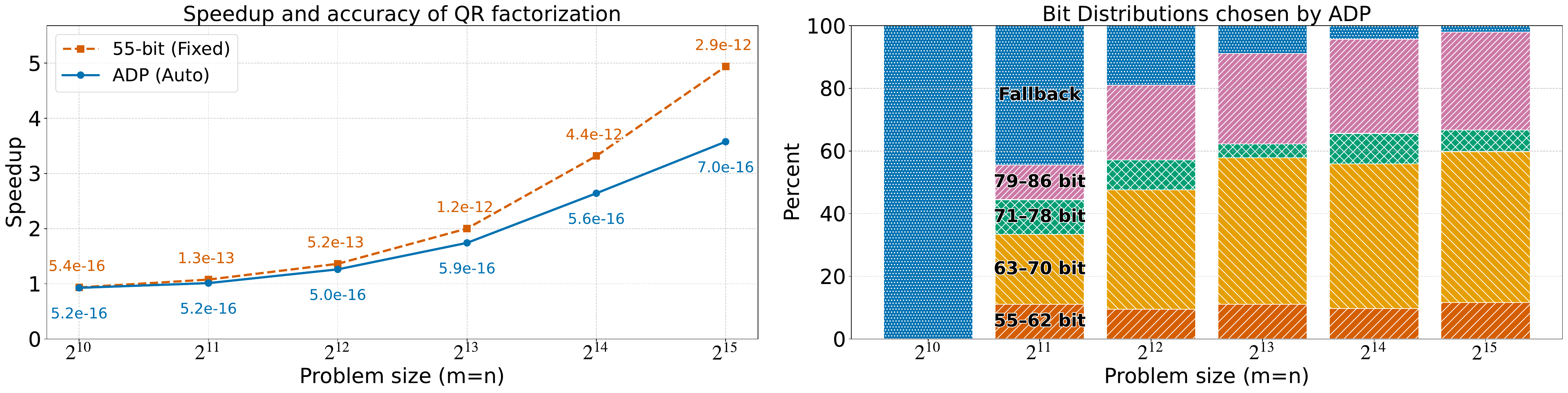}
\Description{Speed up.}
\caption{Application-level evaluation of ADP in QR factorization on RTX Pro 6000 Blackwell Server Edition. 
We modified \texttt{cusolverDnGeqrf} 
so that the trailing matrix updates are performed with emulated DGEMM. 
(Left) End-to-end speedups relative to native FP64 for two configurations: 
fixed emulation at 55 mantissa bits with no ADP (performance upper bound) 
and ADP in dynamic mode, where the number of mantissa bits is selected 
at run time. The absolute residual of the factorization is shown.
(Right) Distribution of slice counts chosen by ADP across all GEMMs in 
the factorization.}
\label{fig:speedup-cusolverdngeqrf}
\end{figure*}

Figure~\ref{fig:speedup-cusolverdngeqrf} presents results for random input matrices 
across a range of sizes on RTX Pro 6000 Blackwell Server Edition.
The left chart reports speedups relative to native FP64 for 
two configurations: (i) fixed emulation with 55 mantissa bits and no ADP 
(upper-bound performance), and (ii) ADP in dynamic mode, where the 
framework selects the number of mantissa bits at run time. The right chart shows
the distribution of slice counts that ADP chooses across all matrix-matrix
multiplications in the factorization.

Several observations emerge. First, ADP achieves end-to-end speedups of 
up to \textbf{3.7$\times$} compared to native FP64, while still delivering 
accuracy comparable to FP64 in the overall factorization. Second, although the 
fixed 55-bit emulation achieves slightly higher speedups, its residuals 
degrade gradually with matrix size, whereas ADP consistently matches the 
accuracy of native FP64 across all problem sizes. Third, the 
performance gap between the fixed 55-bit ceiling and ADP is modest, 
demonstrating that the run time guardrails do not significantly reduce 
throughput. Fourth, the slice distribution in the right chart highlights 
how ADP adapts dynamically: most GEMMs require only a small number of 
slices (e.g., 8–9), while higher slice counts are used only rarely when 
numerical fidelity demands it. Finally, for very small problem sizes, ADP 
recognizes that the overhead of emulation outweighs its benefits and 
therefore falls back to native FP64, ensuring that performance never 
drops below the baseline. Together, these results show that ADP’s 
heuristics are both conservative enough to guarantee accuracy and 
efficient enough to avoid unnecessary cost.

It is worth noting that ADP introduces two types of overhead. The first 
is fixed: run time checks and guardrails that apply to every GEMM call. 
The second is dynamic: in cases where ADP decides to use more slices, the 
extra slicing and integer multiplications proportionally increase cost. 
Despite these factors, the net impact remains favorable: ADP consistently 
delivers large end-to-end speedups while ensuring FP64-level reliability 
under all inputs.

Overall, this experiment demonstrates that ADP is not only effective in 
isolated GEMM benchmarks but also deployable in production solvers. Its 
integration into \texttt{cuSOLVER} shows that ADP can serve as a transparent and 
safe drop-in replacement for native FP64 in real HPC workflows.

\tikzset{
  fc/line/.style     = {->, thick}, 
  fc/block/.style    = {rectangle, rounded corners=2pt, draw, align=center,
                        minimum width=3.8cm, minimum height=1.1cm},
  fc/decision/.style = {diamond, aspect=2, draw, align=center, inner sep=1.5pt},
  fc/term/.style     = {ellipse, draw, align=center, minimum width=2.8cm, minimum height=1.1cm}
}

\usetikzlibrary{positioning,arrows.meta}

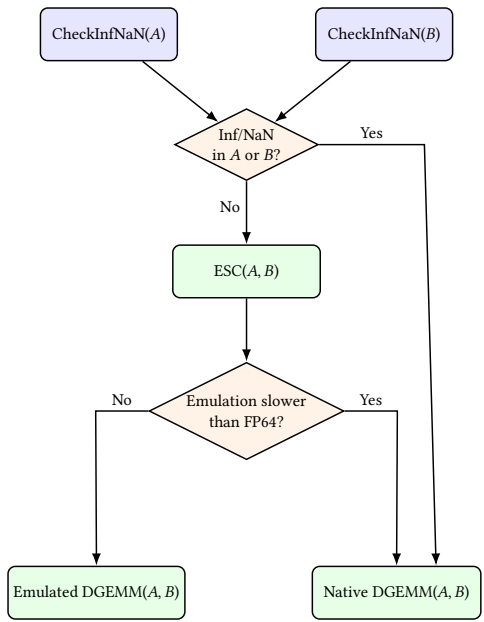
\begin{figure}[ht]
\centering
\scalebox{0.7}{
\begin{tikzpicture}[scale=1.0, >=Latex,thick,font=\normalsize, every node/.style={scale=1.0}]

\tikzstyle{block} = [draw,rounded corners,align=center,fill=blue!10,minimum height=1cm,minimum width=2.6cm]
\tikzstyle{decision} = [draw,diamond,aspect=2,align=center,fill=orange!10,inner sep=1pt]
\tikzstyle{line} = [->,thick]

\node[block] (a) {CheckInfNaN($A$)};
\node[block,right=2.6cm of a] (b) {CheckInfNaN($B$)};

\node[decision,below=1.4cm of $(a)!0.5!(b)$] (decide) {Inf/NaN \\in $A$ or $B$?};

\node[block,below=1.2cm of decide,minimum width=2.8cm,fill=green!10] (esc) {ESC($A,B$)};

\node[decision,below=1.2cm of esc] (hdec) {Emulation slower\\than FP64?};

\node[block,below=2cm of hdec,xshift=-2.85cm,minimum width=3.2cm,fill=green!10] (ozaki) {Emulated DGEMM($A,B$)};
\node[block,below=2cm of hdec,xshift= 2.85cm,minimum width=3.2cm,fill=green!10] (fp64) {Native DGEMM($A,B$)};

\path (fp64.north) ++(0.75,0) coordinate (vfp);

\draw[line] (a) -- (decide);
\draw[line] (b) -- (decide);

\draw[line] (decide.east) -- node[midway,above]{Yes}  ++(2.0,0) -- (vfp);

\draw[line] (decide) -- node[pos=0.4,left]{No} (esc);
\draw[line] (esc) -- (hdec);

\draw[line] (hdec.east) -- node[midway,above]{Yes}  ++(1.0,0) -- (fp64.north);
\draw[line] (hdec.west) -- node[midway,above]{No}  ++(-1.0,0) -- (ozaki.north);

\end{tikzpicture}
}
\caption{Flowchart for GEMM decision process.}
\label{fig:gemm-decision}
\end{figure}

\section{Discussion}
\label{sec:discussion}

The evaluation results confirm that FP64 emulation guided by ESC 
and realized through ADP is both safe and efficient. Taken together, 
our findings show that Ozaki-style schemes—once confined to research 
prototypes—can now be deployed as practical building blocks of 
production HPC libraries. We highlight three main outcomes.

\subsection{Safety through ADP Guardrails}
A central contribution of this work is demonstrating that ADP can 
serve as a lightweight, GPU-resident guardrail for emulation. By 
analyzing exponent spans and detecting exceptional values, ESC 
determines the minimal number of slices needed to guarantee FP64 
accuracy. If the requirement exceeds what is practical or if it encounters any corner cases,
ADP falls back automatically to native FP64. This ensures that users and applications 
are never exposed to silent accuracy loss. The result is that ADP-enabled 
DGEMM can be adopted as a drop-in replacement without requiring 
application developers to reason about slice counts or precision tuning.

\subsection{Efficiency in Practice}
Despite introducing guardrails, ADP incurs modest overhead: less 
than 10\% of run time even under worst-case configurations, and 
smaller still in typical use cases. Our results show speedups of 
up to 2.3$\times$ on GB200 and 13.2$\times$ on RTX Pro 6000 Blackwell Server Edition over 
native FP64, respectively.

These results demonstrate that HPC workloads can be accelerated safely
with minimal friction: existing solvers can benefit from ADP without 
substantial code modifications. As a case study, we modified the QR 
factorization routines in \texttt{cuSOLVER} so that their trailing 
matrix updates were dispatched to ADP-enabled GEMM. This shows how 
ADP can be integrated into production solvers transparently, while 
delivering both performance gains and accuracy guarantees. More 
importantly, once ADP-enabled GEMM is provided in \texttt{cuBLAS}, 
higher-level libraries such as \texttt{cuSOLVER}, \texttt{MAGMA}~\cite{10.1177/10943420241261960}, and \texttt{PETSc}~\cite{osti_2337606}
can inherit these benefits automatically by linkage.
This productization of ADP-enabled DGEMM into \texttt{cuBLAS} underscores the 
confidence in the method: what began as a research technique 
is now positioned as a core technology of NVIDIA’s math libraries.

\subsection{Implications for HPC and AI Hardware}
Our results also carry broader implications. On the software side, 
they suggest a path toward mainstreaming emulated FP64 as a 
first-class execution mode in GPU libraries, rather than a 
research-only alternative. On the hardware side, they illustrate 
that existing low-precision accelerators can be made trustworthy 
for double-precision matrix multiplication oriented workloads, and point toward future designs 
where precision modes are flexibly exposed and managed at 
run time by guardrails such as ESC.

\subsection{Limitations and Future Directions}
While ADP ensures correctness and efficiency, the conservative 
nature of ESC means that its bounds are not always tight. In some 
cases, this leads to overestimation of the number of mantissa bits required. 
The consequence is twofold: first, ADP may incur additional operations 
by using more slices than strictly necessary, and second, in extreme 
cases, the estimate may exceed what is practical, triggering a fallback 
to native FP64. Both outcomes preserve correctness but can reduce 
performance. Developing tighter bounds and refined run time heuristics 
to mitigate overestimation remains an important avenue for future work.

Overall, this discussion underscores that ADP bridges the gap 
between the hardware reality of low-precision accelerators and the 
software demand for reliable FP64. It offers a safe, efficient, and 
transparent foundation for the next generation of numerical libraries 
and HPC applications.

\section{Conclusion}
\label{sec:conclusion}

Modern GPUs deliver their highest performance through low-precision 
Tensor Cores, yet much of scientific computing continues to demand 
FP64 accuracy. Bridging this gap requires techniques that are not 
only fast, but also safe and practical for production use. 

This work introduced the \emph{Exponent Span Capacity (ESC)} estimator 
and the \emph{Automatic Dynamic Precision (ADP)} framework as a path 
to deployable FP64 emulation for matrix multiplication. ESC provides a hardware-agnostic method 
for conservatively estimating the number of slices needed to guarantee 
accuracy, while ADP builds on ESC to integrate run time heuristics, 
exception handling, and automatic fallback to native FP64. Together, 
they ensure that emulated GEMM is both efficient and numerically reliable. 
The accuracy experiments demonstrate that our Ozaki-I DGEMM implementation behaves in every aspect like standard native DGEMM. We further improved Ozaki-style decomposition with an unsigned slicing 
representation that increases bit utilization, reducing slice counts 
and improving throughput.

Our evaluation shows that ADP overhead remains below 10\% even under 
worst-case configurations, and that emulated GEMM achieves up to 
2.3$\times$ speedups on GB200 and 13.2$\times$ on RTX Pro 6000 Blackwell Server Edition compared 
to native FP64. Application-level experiments with QR factorization 
demonstrate that ADP-enabled GEMM implementations can be integrated into HPC workloads with minimal 
friction, achieving up to 3.7$\times$ end-to-end speedups while 
maintaining accuracy on par with FP64. Importantly, once ADP-enabled 
GEMM is adopted in \texttt{cuBLAS}, higher-level libraries and frameworks
will benefit automatically through 
linkage, seamlessly into scientific computing applications.

While ADP guarantees correctness, the conservative nature of ESC means 
that its bounds are not always tight. Overestimation of required slices 
can lead to unnecessary overhead or, in extreme cases, fallback to native 
FP64. Both outcomes preserve accuracy but reduce performance. Tightening 
ESC’s estimates and refining heuristics to minimize overestimation remain 
important directions for future work. Extending ADP beyond dense GEMM
also represents a promising path forward.

Although our evaluation focused on Ozaki Scheme I, the techniques 
introduced in this paper are general and can be extended to other 
emulation schemes such as Ozaki Scheme II, underscoring the broader 
applicability of ESC and ADP beyond the specific instantiation studied here.
Additionally, it is straightforward to extend the emulation of DGEMM, including the APD framework, to ZGEMM via the 4M method~\cite{10.1145/3086466}. 

Implementations of ESC and our Ozaki-I emulation framework will be made 
available as open source, enabling the community to build upon this work 
and integrate these techniques into HPC libraries and applications.

In conclusion, ADP and ESC transform FP64 matrix multiplication emulation from a research 
prototype into a safe deployable library implementation. They enable low-precision 
accelerators—originally designed for AI—to serve as trustworthy building 
blocks for high-fidelity HPC workloads. This work demonstrates that 
scientific computing can reap the performance benefits of AI-driven 
hardware without compromising numerical accuracy.


\bibliographystyle{ACM-Reference-Format}   
\bibliography{bibliography} 

\end{document}